\documentclass[12pt]{article}

\pdfoutput=1

\usepackage{a4wide}
\usepackage{amssymb}
\usepackage{graphicx}
\usepackage{verbatim}
\usepackage{textcomp}

\usepackage[intlimits]{amsmath}

\usepackage{amsfonts}

\usepackage{epsfig}

\newcommand{\be}{\begin{equation}}
\newcommand{\ee}{\end{equation}}
\newcommand{\bea}{\begin{eqnarray}}
\newcommand{\eea}{\end{eqnarray}}
\newcommand{\ba}{\begin{eqnarray}}
\newcommand{\ea}{\end{eqnarray}}

\def\XXint#1#2#3{{\setbox0=\hbox{$#1{#2#3}{\int}$}
     \vcenter{\hbox{$#2#3$}}\kern-.52\wd0}}

\begin{document}

\setcounter{table}{0}

\begin{flushright}\footnotesize

\texttt{ICCUB-20-022}

\end{flushright}

\mbox{}
\vspace{0truecm}
\linespread{1.1}

\vspace{0.5truecm}

\centerline{\Large \bf Phases of  unitary matrix models and lattice QCD2} 


\vspace{1.3truecm}

\centerline{
    {\large \bf Jorge G. Russo} }

\vspace{0.8cm}

\noindent  
\centerline {\it Instituci\'o Catalana de Recerca i Estudis Avan\c{c}ats (ICREA), }
\centerline{\it Pg. Lluis Companys, 23, 08010 Barcelona, Spain.}

\medskip
\noindent 
\centerline{\it  Departament de F\' \i sica Cu\' antica i Astrof\'\i sica and Institut de Ci\`encies del Cosmos,}
\centerline{\it Universitat de Barcelona, Mart\'i Franqu\`es, 1, 08028 Barcelona, Spain. }

\medskip

\centerline{  {\it E-Mail:}  {\texttt jorge.russo@icrea.cat} }

\vspace{1.2cm}

\centerline{\bf ABSTRACT}
\medskip

We investigate the different large $N$ phases of  a generalized Gross-Witten-Wadia $U(N)$ matrix model. The deformation mimics the one-loop determinant of fermion matter with a particular coupling to gauge fields.
In one version of the model, the GWW phase transition is smoothed out and it becomes a crossover. In another version, the phase transition occurs along a critical line in
the two-dimensional parameter space spanned by the 't~Hooft coupling $\lambda$ and the Veneziano parameter $\tau$. We compute the expectation value of Wilson loops in both phases, showing that the transition is third-order. A calculation of the $\beta $
function shows the existence of an IR stable fixed point.

\noindent

\vskip 1.2cm
\newpage

\def\sech{ {\rm sech}}
\def\p{\partial}
\def\pa{\partial}
\def\ov{\over }
\def\a{\alpha }
\def\g{\gamma}
\def\s{\sigma }
\def\td{\tilde }
\def\vp{\varphi}
\def\strokedint{\int}
\def \ha {{1 \over 2}}

\def\KK{{\cal K}}




\textwidth = 460pt
\hoffset=0pt



\section{Introduction}

 The study of random matrix ensembles     caters to a broad variety of most diverse applications  in different areas of physics and mathematics \cite{mehta,forrester,baik}.
In a recent paper \cite{Russo:2020eif} a new unitary  one-matrix model was constructed and investigated. The model has the potential
\begin{equation}
    V=-\frac{1}{g^2} \, {\rm Tr}\left(U+U^\dagger\right) + 2\nu \, 
   {\rm Tr} \ln \left( 2+U+U^\dagger \right)\ .
\label{uno}
\end{equation}
It represents a natural deformation of the Gross-Witten-Wadia (GWW)
model \cite{Gross:1980he,Wadia:2012fr} and it interpolates between classical unitary matrix models. 
The theory exhibits various large $N$ phase transitions in the parameter
space of the two couplings, with an intricate structure that has been only partially investigated in  \cite{Russo:2020eif},
and it is the unitary counterpart of the Hermitian, deformed-Cauchy random matrix ensemble \cite{Russo:2020pnv}. This contains one hermitian matrix $M$ subject to the  potential
\begin{equation}
    V_H = A \, {\rm Tr} \ln(1+M^2) +B\, {\rm Tr}\frac{1}{1+M^2}\ .
\label{dos}
\end{equation}
The term with coefficient $A$ corresponds to the logarithmic term in \eqref{uno}
(taking into account a shift due to the Jacobian of the sterographic map),
while the term with coefficient $B$ gives rise to the GWW  term.
The  model  generalizes the Hermitian model appeared in \cite{Santilli:2020ueh}, derived from the $\nu=0$ case (see \cite{Minahan:1991pv,Minahan:1991sm} for earlier studies on related models).
The logarithmic term appears in the mathematical literature in the study of Cauchy random matrix ensembles \cite{WitteForr}.
Above a certain critical value of the coupling $B$, the potential develops a double well,  leading to a phase transition.
In this paper we will describe in detail the analogous phase transition  in the unitary model \eqref{uno}, which exhibits some striking features.
One phase was already described in \cite{Russo:2020eif}. Here we will find the explicit solution for both phases and compute some relevant observables.

The logarithmic term with coefficient $\nu$ corresponds to an insertion
\begin{equation}
  \left[  \det \left(2+U+U^\dagger \right)\right]^\nu=\left[  \det \left(1+U\right)
  \left(1+U^\dagger \right)\right]^\nu\ .
\label{tres}
\end{equation}
A related deformation is obtained by the insertion of the operator:
\begin{equation}
  \left[  \det \left(2-U-U^\dagger \right)\right]^\nu=\left[  \det \left(1-U\right)
  \left(1-U^\dagger \right)\right]^\nu\ .
\label{cuatro}
\end{equation}
This is equivalent to the  deformation \eqref{tres}
if at the same time the sign of the coupling is flipped, $g^2\to -g^2$ and $U\to -U$. 
However, when viewed as deformations of the GWW matrix model, the physical interpretation of both models is different. In particular, the insertion \eqref{tres}
leads to smoothing out the GWW phase transition occurring in the physical range of the coupling $g^2>0$, whereas 
the insertion \eqref{cuatro} leads to extending
 the GWW third-order phase transition to a critical line in the two-dimensional parameter space of couplings. In addition, $U\to -U$ flips the sign of the Wilson loop.

 An important question concerns the possible physical interpretation of the 
deformation in the context of two-dimensional lattice gauge theory.
The determinantal form of the insertion, with $\nu>0$, suggests the obvious interpretation as a one-loop fermion determinant.
In conventional lattice QCD2, one has the Wilson fermions with action \cite{Huang:1988br}
\be
S=\sum_{n,m} \bar \psi_n K(n,m) \psi_m\ ,
\ee
\be
K(n,m)=\delta_{nm} - \frac{1}{2(2r+m_0)} \sum_\mu \big[ (r-\gamma_\mu)U_\mu(n)\delta_{n+\mu,m}+(r+\gamma_\mu)U_\mu^\dagger (m)\delta_{n-\mu,m}\big]\ ,
\label{qcddos}
\ee
where $r$ contributes to the energy of spurious fermion modes and $m_0$ represents the bare quark mass. The fermion determinant is complicated and finding the meson spectrum requires elaborated calculations based on Monte Carlo  \cite{Huang:1988br}.
On the other hand,  the insertion \eqref{tres} is equivalent to
\be
\int D\psi D\bar\psi \, \, e^{\sum_{i=1}^{N_f}\bar\psi_i  \left( 2+ U+U^\dagger \right)\psi_i }\ ,\qquad N_f\equiv \nu/2\ .
\label{bifer}
\ee
The determinant \eqref{tres} 
is clearly much simpler than the one in the more realistic model
\eqref{qcddos}, as it only includes variables in a single plaquette: like in
the GWW matrix model, the full partition function would be obtained as $Z^{V/a^2}$,
where  $V$ is the volume of space and $a$ is the lattice spacing.
However,  similar (but  still more complicated) determinants appear in Eguchi-Kawai reductions  \cite{Eguchi:1982nm}  involving Wilson fermion determinants \cite{Neuberger:2002bk,Kovtun:2007py}. In particular, for $N_f$ adjoint fermions, the adjoint Eguchi-Kawai model has a single-site Wilson fermion operator $D_W$
given by \cite{Bringoltz:2011by}
\be
D_W=1-\kappa\sum_{i=1}^4 \left[(1-\gamma_\mu) U_\mu^{\rm adj} +(1+\gamma_\mu)U_\mu^{\dagger {\rm adj}}\right]\ .
\ee
Computing physical quantities in this model, such as Wilson loop expectation values,
still requires heavy numerical  calculations \cite{Bringoltz:2011by}.
On the other hand, the present model  can be fully solved by analytic methods, including $1/N$ effects, and  it might  provide a simple phenomenological setup to reproduce some  features of QCD2.

A different strategy is to write the bilinear fermion term in \eqref{bifer} as
a gauge-invariant term in the form
\be
\bar\psi_i  \left( 1+\prod_P U+h.c. \right)\psi_i \ ,
\ee
where the product $\prod_P$ represents the square plaquette containing the original  link variables prior to the Weyl gauge fixing $U_{\vec n, \vec \i_0}=1$, $\vec\i_0$ being the lattice vector in the time direction.
In the continuum, the plaquette is associated with the operator ${\rm Tr}\, e^{i a^2 \hat F_{01} }$ (see {\it e.g.} \cite{creutz}).
This naturally leads to an expansion in the bilinear fermion term having higher-dimensional couplings such as
$\bar\psi \hat F_{\mu\nu}\hat F^{\mu\nu} \psi $.
Thus, according to this view, the insertion \eqref{tres} would seem to compute the effect of such deformations together with a mass term, in sectors where the fermion kinetic energy may be negligible.

Another interesting interpretation of the deformation in \eqref{uno} arises in the context of  $SU(N)$ Yang-Mills theory on $S^3$ \cite{Sundborg:1999ue,Aharony:2003sx}. 
Expanding the new term in powers of  $U, \, U^\dagger$, one has
\be
{\rm Tr} \ln \left( 2+U+U^\dagger \right)=- \sum_{k=1}^\infty \frac{(-1)^k}{k} \left( {\rm Tr}\ U^k+ {\rm Tr}\ {U^\dagger}^ k\right)\ .
\ee
This operator plays a special role in the blackhole/string phase transition
as it represents a gap opening perturbation added to the action \cite{AlvarezGaume:2006jg}. It would be extremely interesting to explore the
consequences of our results in that context.

This paper is organized as follows. In section 2 we introduce the matrix models A and B corresponding to insertions \eqref{tres} and \eqref{cuatro}
and study them in a large $N$ double-scaling limit. The eigenvalue densities
for models A and B in the two different phases are determined in \textsection 2.1 and \textsection 2.2.
The resulting phase diagram  is shown in fig. \ref{dosfases}.
In section 3 we compute the free energy, the vacuum 
expectation value of  Wilson loops and winding Wilson loops in the weak and strong coupling phases of model B and determine the order of the phase transition.
Finally, in section 4 we compute the $\beta $ function. In model B it exhibits the presence of IR stable fixed points.

\section{The unitary matrix model}

We shall consider two deformations of  lattice $U(N)$ gauge theory in two dimensions.
The first model A has partition function
\be
Z^A=\int dU\,  \det \left( \frac14 \left(2+U+U^\dagger \right)\right)^\nu e^{\frac{1}{g^2}{\rm Tr}(U+U^\dagger)}\ .
\ee
Integrating over the volume of the group, this becomes
\begin{equation}
Z^A=\frac{1}{N!}\int_{(0,2\pi ]^{N}}\prod_{1\leq j<k\leq N}\left\vert e^{i\varphi
_{j}}-e^{i\varphi _{k}}\right\vert ^{2}\prod_{j=1}^{N}\cos ^{2\nu}\left( 
\frac{\varphi _{j}}{2}\right) \exp \left( \frac{2}{g^2}\cos \left( \varphi
_{j}\right) \right) \frac{d\varphi _{j}}{2\pi}\ .
\label{dosdos}
\end{equation}%
The second model B has partition function
\be
Z^B=\int dU\,  \det \left( \frac14 \left(2-U-U^\dagger \right)\right)^\nu e^{\frac{1}{g^2}{\rm Tr}(U+U^\dagger)}\ .
\ee
In this case, one gets
\begin{equation}
Z^B=\frac{1}{N!}\int_{(0,2\pi ]^{N}}\prod_{1\leq j<k\leq N}\left\vert e^{i\varphi
_{j}}-e^{i\varphi _{k}}\right\vert ^{2}\prod_{j=1}^{N}\sin ^{2\nu}\left( 
\frac{\varphi _{j}}{2}\right) \exp \left( \frac{2}{g^2}\cos \left( \varphi
_{j}\right) \right) \frac{d\varphi _{j}}{2\pi}\ .
\label{dossin}
\end{equation}%
The two models are mathematically equivalent as they are related by changing
$g^2\to -g^2$ and shifting the $\varphi_j$ integration variables by $\pi $.
However, their physical interpretation as a deformation of the GWW model is different.
Considering  $g^2>0$, the deformation of the potential in terms of
$\ln\sin^2\frac{\varphi}{2}$ --instead of $\ln\cos^2\frac{\varphi}{2}$ -- implies a stronger
deformation in the region of small eigenvalues. As a result, this will lead to a very different deviation from  the GWW model: in model A, the large $N$  GWW phase transition is smoothed out and it becomes a crossover \cite{Russo:2020eif}. In model B, the large $N$ GWW phase transition subsists. In addition, there are some striking consequences for  observables and
for the $\beta $ function of the running coupling.

We are interested in the large $N$ ``Veneziano" limit with fixed parameters
\be
\lambda \equiv g^2N\ ,\qquad \tau \equiv\frac{\nu}{N}\ .
\ee

Thus we have  unitary one-matrix models with potentials
\bea
V^A &=&-\frac{2}{\lambda}\, \cos(\alpha )-\tau \ln \cos^2\frac{\alpha}{2}\ ,
\label{potA}\\
V^B &=&-\frac{2}{\lambda}\, \cos(\alpha )-\tau \ln \sin^2\frac{\alpha}{2}\ .
\label{potB}
\eea
Here we assume that $\tau\geq 0$ (a discussion of the phase diagram in
the region $\tau<0$ is given in \cite{Russo:2020eif}).
A partial analysis was carried out in \cite{Russo:2020eif}, where it was found that these models have two phases. The phase transition occurs on a critical line
$\lambda_{\rm cr}(\tau )$, which for model B lies on the region $\lambda> 0$.
As only one phase was described explicitly in \cite{Russo:2020eif}, here we will complete this analysis
by explicitly deriving the solution in the two phases and by computing the free energy and some relevant physical observables.

\subsection{Eigenvalue distribution for model A}

At large $N$, the partition function is determined by a saddle-point calculation.
The large $N$ regime is studied as usual by introducing a unit-normalized
density of eigenvalues $\rho(\alpha)$. The saddle-point equation then becomes the singular integral
equation
\begin{equation}
\frac{2}{\lambda}\,\sin\alpha + \tau\tan\frac{\alpha}{2} =P\int_{L}
d\beta\, \rho(\beta )\cot\left( \frac{\alpha-\beta}{2}\right)\ ,
\label{ecurho}
\end{equation}
where $L$ represents the region where eigenvalues condense.
The dynamics governing the eigenvalues can be understood by examining
the behavior of the potential in the parameter space. 

Consider, in first place, positive $\lambda$.
In this case both terms of the potential \eqref{potA} give a force driving  eigenvalues to the region near $\alpha=0$. For small $\lambda $, the force is large and eigenvalues must get condensed in a small cut, with an  eigenvalue distribution that must approach the GWW eigenvalue distribution, since the deformation is negligible 
compared with the first term.
 As $\lambda $ is increased, the cut gets wider. However, as long as $\tau> 0$, eigenvalues  cannot get to $\pm \pi$ because the potential grows to infinity at $\pm \pi $ owing to the deformation. This is
a crucial effect, which removes the GWW phase transition transforming it into a crossover.

On the other hand, for negative $\lambda $, the force associated with the sine term
in \eqref{ecurho} becomes repulsive. As a result, there is a critical coupling beyond which 
 the potential develops a double well. This occurs at $\lambda_1=-4/\tau$.
At this point, the eigenvalue distribution is still  described by a one-cut solution, due to overfilling of eigenvalues (a fact that will be verified below).
By further increasing $\lambda $, one meets another critical value $\lambda_{\rm cr}$, where the
eigenvalue distribution is split into two cuts.
Below we explicitly describe the two regimes.

\subsubsection{The one-cut phase in model A}

Let us first review the main features of the one-cut solution found in
 \cite{Russo:2020eif}. This phase is described by the solution
\begin{equation}
    \rho(\alpha ) =\left( \frac{2}{\pi\lambda} \cos\frac{\alpha}{2} +\frac{\tau}{2\pi}\frac{1}{\sqrt{1-m}\cos\frac{\alpha}{2}}
\right) \sqrt{m-\sin^2\frac{\alpha}{2}}\ .
\label{soluno}
\end{equation}
The cut extends in the interval $(-\alpha_0,\alpha_0)$, with $m=\sin^2\alpha_0/2$,
$|\alpha_0|<\pi$, $0<m<1$.
When $\tau=0$, the solution \eqref{soluno} reduces to the familiar solution of the GWW model in the gapped phase.
The parameter $m$ determines
the width of the eigenvalue distribution.
It is easily found from the normalization condition,
\begin{equation}
1=\int_{-\alpha_0}^{\alpha_0}
d\beta\, \rho(\beta )=\frac{2m}{\lambda}
+\left(\frac{1}{\sqrt{1-m}}-1\right) \tau\ .
\label{normales}
\end{equation}
which leads to a cubic equation for $m$. For all $\tau>0$, there is a unique real root with 
$0<m<1$, which  determines $m=m(\lambda,\tau)$. A simple way to see this is by solving the normalization condition for $\tau $. This gives 
\begin{equation}
\tau = \frac{\sqrt{1-m}\ (\lambda -2 m)}{\lambda  \left(1-\sqrt{1-m}\right)}\ .
\label{taunorma}
\end{equation}
For small $m>0$ one has $\tau\sim 2/m\gg 1$. Then $\tau $ monotonically decreases
until it vanishes. Thus for any positive $\tau $ there is a unique value of $m$ satisfying \eqref{taunorma}.

As discussed above, the phase described by \eqref{soluno} subsists until 
a critical point beyond which  $\rho$ becomes negative in some  interval
 $\subset (-\alpha_0,\alpha_0)$, due to the fact that the potential develops a double well.
At the critical point the density $\rho $ vanishes at the origin $\alpha=0$. This gives the condition
\be 
\frac{2}{\pi\lambda}  +\frac{\tau}{2\pi}\frac{1}{\sqrt{1-m}}=0\ \longrightarrow\ \tau \lambda_{\rm cr} =-4\sqrt{1-m}\ .
\ee
In particular, this shows that $0>\lambda_{\rm cr}>\lambda_1$. Since $m$ is a function of $\lambda $ and $\tau$, the  critical coupling $\lambda_{\rm cr}$ is a function of $\tau $. Indeed, combining with \eqref{taunorma}, we obtain
\be
\lambda_{\rm cr} = -\frac{4}{\tau ^2}\left(\tau+1 -\sqrt{2 \tau +1}\right)\ ,
\label{lambdaci}
\ee
or
\be
\frac{2}{\lambda_{\rm cr}} =-\frac12 \left(\tau+1+\sqrt{2\tau+1}\right)\ .
\ee
Thus the eigenvalue density  is described by the solution \eqref{soluno} in the 
regime $\lambda \in \{-\infty,\lambda_{\rm cr}\} \cup \{0,\infty\}$.

Note that, for  $\tau\to 0 $, one has $\lambda_{\rm cr}\to -2$.
This is nothing but the GWW phase transition in a frame where 
 the sign of $\lambda $ has been flipped by shifting the eigenvalues by $\pi $.
On the other hand, the GWW phase transition occurring at $\tau=0$, $\lambda=2$ is smoothed out by the deformation. For any $\tau>0$, the free energy in the region $\lambda>0$ is analytic, since the solution does not change in this region 
of the plane $(\tau,\lambda)$. Of course,
 the phase transition at $\lambda=2$ remains on the axis $\tau=0$.


\subsubsection{The two-cut phase in model A}

In the regime $0>\lambda >\lambda _{\rm cr}$, the eigenvalue distribution is split into
two cuts, implying a phase transition.
The phase transition is the unitary matrix model counterpart of the phase transition found in \cite{Russo:2020pnv}, in the Hermitian matrix model
version.
The solutions of  unitary and Hermitian matrix models are connected
by a simple map described in
\cite{Mizoguchi:2004ne} (see also appendix in \cite{Okuyama:2017pil}).
Alternatively, one can solve \eqref{ecurho} directly by standard methods
for one-matrix models.
For the two-cut solution we obtain

\be
\rho(\alpha ) =  \frac{\tau}{4\pi} \, \sqrt{ \sin^2 \alpha }
\sqrt{a^2-\tan^2 \frac{\alpha}{2} } \sqrt{\tan^2 \frac{\alpha}{2} -b^2 } \ .
\label{solurho2}
\ee
The parameters $a$, $b$ representing the endpoints of the eigenvalue distribution
may be obtained from two conditions arising from normalization and from
the integral equation \eqref{ecurho} itself.
To compute integrals and check the equations, it is convenient
to introduce the variable $t=e^{i\alpha}$.
Then
\be
\rho(\alpha) d\alpha = \hat \rho(t) dt\ ,
\ee
with
\be
\hat \rho(t) =\frac{\tau}{8\pi}\, \frac{(1-t)}{(1+t) t^2}\ \sqrt{4t-(1+b^2)(1+t)^2}
\sqrt{a^2(1+t)^2+(t-1)^2}\ .
\ee
We  first demand normalization:
\be
\int_L dt \hat \rho(t) =1\ .
\ee
The integral goes over the two cuts in the circle described by $t$, 
$L= (\alpha_0,\alpha_1)\cup (-\alpha_1,-\alpha_0)$.
The integral can be computed by residues, by considering two contours, surrounding 
each branch cut. Then the integral picks the residue of the poles at $t=0$, $t=-1$ and $t=\infty $,
that is
\be
\int_L dt \hat \rho(t) = i\pi \left( {\rm Res}_{t=0} \hat \rho(t)+{\rm Res}_{t=-1} \hat \rho(t)+{\rm Res}_{t=\infty}  \hat \rho(t)\right)\ .
\ee
We obtain the condition
\be
1=\tau \left(\frac{2+a^2+b^2}{2\sqrt{1+a^2}\sqrt{1+b^2} } -1 \right)\ .
\label{normab}
\ee
Let us now consider the integral equation \eqref{ecurho}. In terms of $t$, it is given by
\be
\frac{2}{\lambda}\sin\alpha +\tau \tan\frac{\alpha}{2} =
-i\int_L dt \hat \rho(t)\ \frac{t+e^{i\alpha}}{t-e^{i\alpha}}\ .
\ee
Choosing the same contours surrounding the cuts, the integration by residues now
gives
\be
-i\int_L dt \hat \rho(t)\ \frac{t+e^{i\alpha}}{t-e^{i\alpha}}=
-\frac{\tau}{2}\sqrt{1+a^2}\sqrt{1+b^2}\, \sin\alpha +\tau \tan\frac{\alpha}{2} \ .
\ee
Therefore we get the second condition on the parameters $a,b$:
\be
\frac{2}{\lambda}= -\frac{\tau}{2}\sqrt{1+a^2}\sqrt{1+b^2}\ .
\label{equab}
\ee
The solution to \eqref{normab}, \eqref{equab} is
\be
1+a^2=  -\frac{4 \left(1+\tau +\sqrt{2 \tau +1}\right)}{\lambda  \tau ^2}\ ,
\ee
\be
1+b^2=  -\frac{4 \left(1+\tau -\sqrt{2 \tau +1}\right)}{\lambda  \tau ^2}\ .
\ee
These values of $a$, $b$ coincide, as expected, with the values of $a$, $b$ of the Hermitian model in \cite{Russo:2020pnv}, taking into account the shift $\tau\to\tau-1$.\footnote{The origin of this shift was explained in  \cite{Russo:2020eif}. It arises from a contribution from the Jacobian
of the transformation in going from the real line to the unit circle.}

The critical line occurs when $b=0$.
This gives
\be
\lambda_{\rm cr}(\tau) =-\frac{4}{\tau ^2} \left(\tau+1 -\sqrt{2 \tau +1}\right)\ .
\ee
This critical line exactly coincides with the critical line obtained from the one-cut phase.
Moreover, we can check continuity on the critical line.
Setting $\lambda =\lambda_{\rm cr}(\tau)$, on the critical line the eigenvalue density \eqref{solurho2}
simplifies to
\be
\rho_{\rm cr} = \frac{\tau}{2\pi }\, \frac{\sin^2 \frac{\alpha}{2}}
{\cos \frac {\alpha}{2}}\, \sqrt{a^2-(1+a^2)\sin^2\frac{\alpha}{2}}\ .
\ee
This  matches the critical density obtained from the solution in phase 1,
noticing that
\be
m_{\rm cr}=\frac{a^2}{1+a^2}\ ,\qquad \lambda_{\rm cr}\tau=-4\sqrt{1-m_{\rm cr}}\ .
\ee



\subsection{Eigenvalue distribution for model B}

We now consider the matrix model defined by the GWW partition function
with the  insertion 
$\det \left( \frac14 \left(2-U-U^\dagger \right)\right)^\nu$,
leading to \eqref{dossin}.
At large $N$, the saddle-point equations are equivalent to the 
integral equation
\begin{equation}
\frac{2}{\lambda}\,\sin\alpha - \tau\cot\frac{\alpha}{2} =P\int_{L}
d\beta\, \rho(\beta )\cot\left( \frac{\alpha-\beta}{2}\right)\ .
\label{ecurhoB}
\end{equation}
Using the connection to model A by $\lambda\to -\lambda$, one finds
that the model B \eqref{dossin} has two phases in the region $\lambda>0$,
separated by a critical line
\be
\lambda_{\rm cr} = \frac{4}{\tau ^2}\left(\tau+1 -\sqrt{2 \tau +1}\right)\ .
\label{lambdacriplus}
\ee
Note that $\lambda_{\rm cr}\to 2$ when $\tau\to 0$.

\begin{figure}[h!]
 \centering
 \begin{tabular}{cc}
 \includegraphics[width=0.4\textwidth]{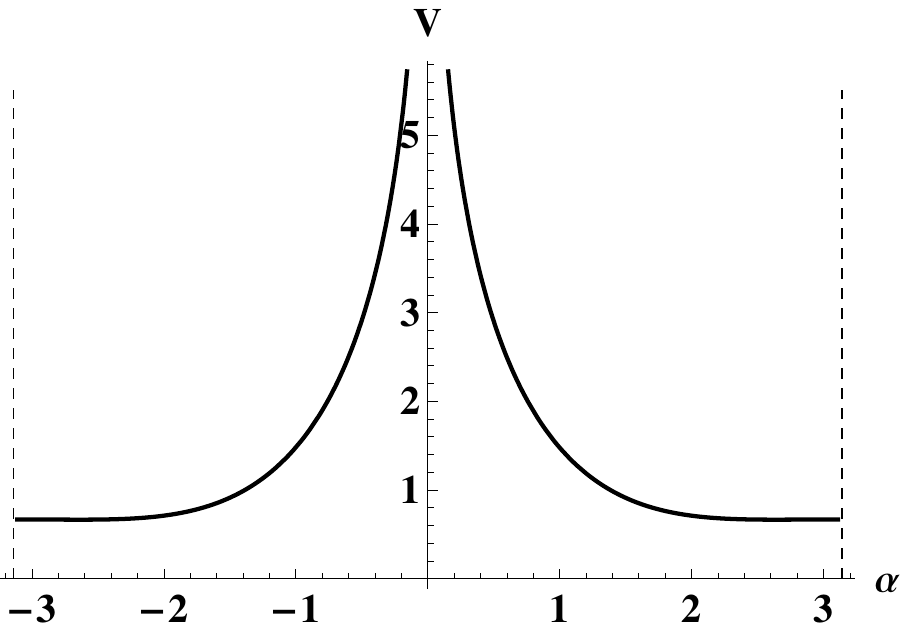}
 &
 \qquad \includegraphics[width=0.4\textwidth]{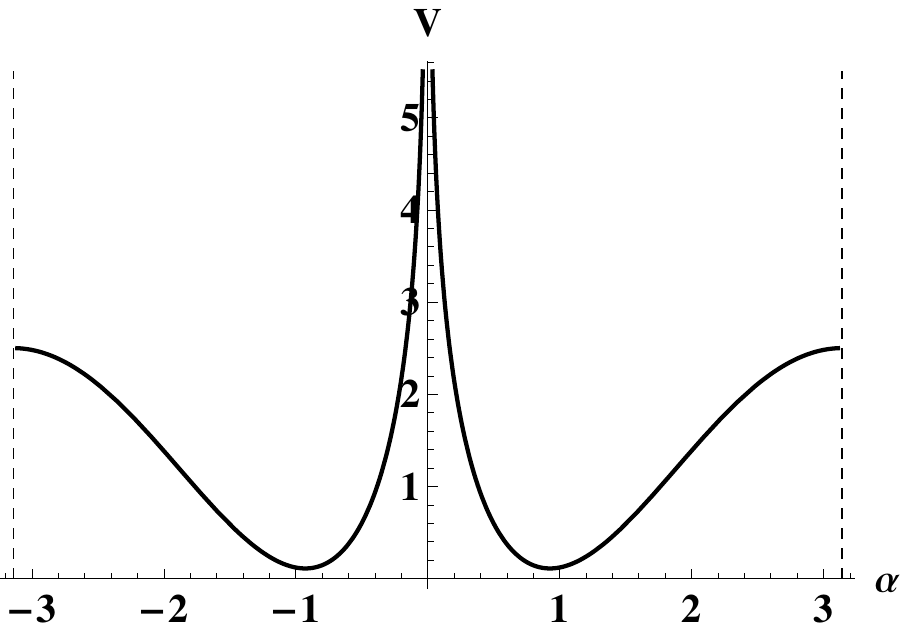}
 \\ (a)&(b)
 \end{tabular}
 \caption
 { The potential in model B. (a) In the supercritical phase, with 
 $\lambda =3$, $\tau =1.25 $.
 (b) In the subcritical phase, with $\lambda =0.8$, $\tau =1$. 
 }
 \label{potencB}
 \end{figure}

The origin of the two phases can be understood by looking at the potential, which has
a double-well for sufficiently small $\lambda$, thus inducing the phase transition from a one-cut to a two cut distribution (see fig. \ref{potencB}). There is a sharp
distinction with the potential in the GWW model. In model B a ``wall" appears at {\it small} eigenvalues, which becomes infinitely thin as $\tau\to 0$, where the physics of the GWW model is recovered.
Below we describe the eigenvalue densities for the two phases.

\subsubsection{One-cut phase in model B ($\lambda>\lambda_{\rm cr}$) }

In the strong coupling phase $\lambda>\lambda_{\rm cr}$  eigenvalues
condense in one cut. The density is obtained from \eqref{soluno}
by using the map $\lambda\to -\lambda$, $\alpha\to \alpha \pm \pi $. One obtains\footnote{In \eqref{soluno}, $\cos \alpha/2$ carries absolute value bars, which can be omitted in the interval $(-\pi,\pi)$. Outside this interval, the density is periodic with period $2\pi $. }
\be
 \rho_{\rm 1\, cut}  (\alpha) =\left(-\frac{2}{\pi  \lambda}\, |\sin\frac{\alpha}{2}|+\frac{\tau}{2\pi}\frac{1}{\sqrt{1-m} \, \,  |\sin\frac{\alpha}{2}|}\right)\, \sqrt{m-\cos^2\frac{\alpha}{2}}\ ,
 \label{rhosenos}
\ee
where $\alpha \in (-\pi,-\alpha_1)\cup (\alpha_1,\pi)$, $\alpha_1=2\arccos\sqrt{m}$.
Now the normalization condition gives
\begin{equation}
1=\int_{-\alpha_0}^{\alpha_0}
d\beta\, \rho_{\rm 1\, cut}(\beta )= -\frac{2m}{\lambda}
+\left(\frac{1}{\sqrt{1-m}}-1\right) \tau\ .
\label{normalsin}
\end{equation}
This uniquely determines $m$ in the interval $0<m<1$ for any $\lambda>0$, $\tau>0$.
Clearly, this is the same as the density \eqref{soluno}, with the argument
shifted by $\pi $
and $\lambda\to -\lambda$.
The critical line occurs when $\rho(\pm \pi )=0$. This gives the condition
$4\sqrt{1-m}=\tau \lambda $, which, combined with \eqref{normalsin}, leads to 
\eqref{lambdacriplus}.

\begin{figure}[h!]
 \centering
 \begin{tabular}{cc}
 \includegraphics[width=0.4\textwidth]{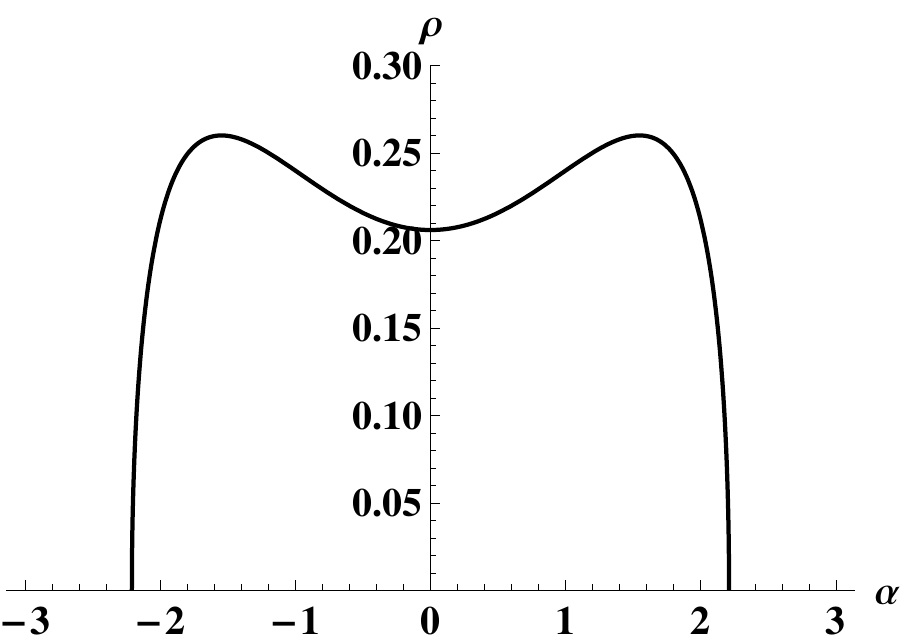}
 &
 \qquad \includegraphics[width=0.4\textwidth]{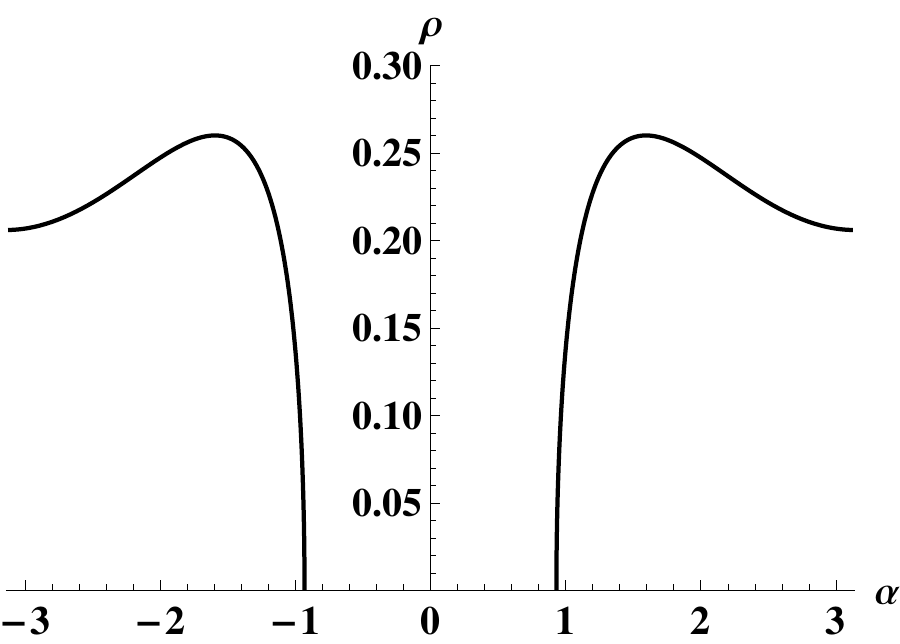}
 \\ (a)&(b)
 \end{tabular}
 \caption
 {(a) Eigenvalue density in model A for $\lambda =-3$, $\tau =1.25 $.
 (b) Eigenvalue density in model B for $\lambda =3$, $\tau =1.25$. It differs
 from the density of fig. (a) by a shift of $\pi$.
 }
 \label{densitysincos}
 \end{figure}

In the infinite coupling limit, the eigenvalue density assumes the asymptotic form
\be
 \rho_{\rm 1\, cut}  (\alpha)\bigg|_{\lambda=\infty} =\frac{\tau}{2\pi}\frac{1}{\sqrt{1-m_\infty} \, |\sin\frac{\alpha}{2}|}\, \sqrt{m_\infty-\cos^2\frac{\alpha}{2}}\ ,\qquad m_\infty =\frac{1+2\tau}{(1+\tau)^2}\ .
 \label{rhoinfinity}
\ee

\subsubsection{Two-cut phase in model B ($0<\lambda<\lambda_{\rm cr}$) }

\begin{figure}[h!]
 \centering
 \includegraphics[width=0.45\textwidth]{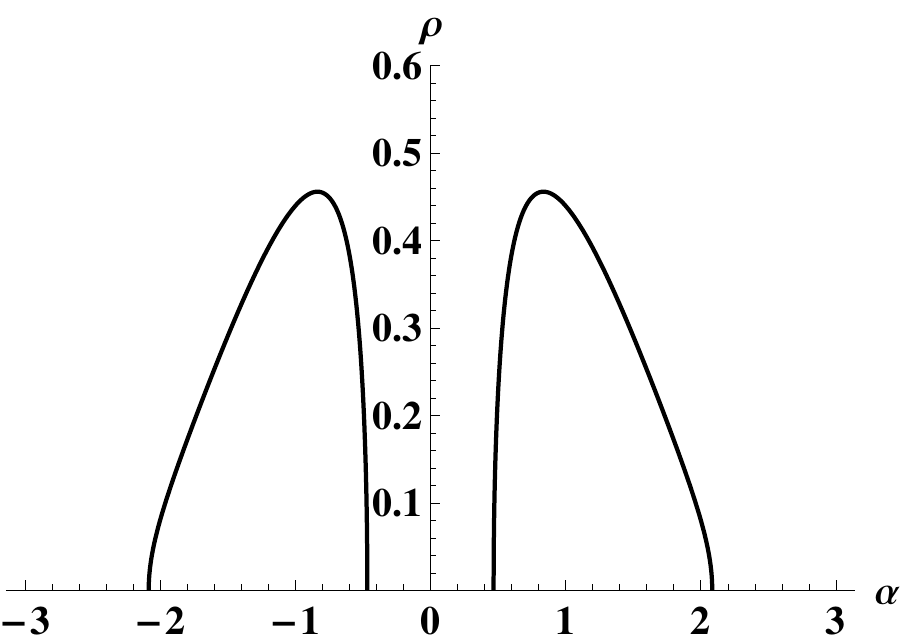}
 \caption{Eigenvalue density in the two-cut (weak coupling) phase in model B. Here $\tau=1$, $\lambda=0.8$.}
 \label{eigentwo}
 \end{figure}

In the weak coupling phase the eigenvalue density has support on two cuts.
Applying the map $\lambda\to -\lambda$, $\alpha\to \alpha \pm \pi $ to \eqref{solurho2}, we get
\be
 \rho_{\rm 2\, cuts}(\alpha ) =  \frac{\tau}{4\pi} \, \sqrt{ \sin^2 \alpha }\, 
\sqrt{a^2-\cot^2 \frac{\alpha}{2} } \sqrt{\cot^2 \frac{\alpha}{2} -b^2 } 
\ ,\qquad \lambda<\lambda_{\rm cr}(\tau)\ ,
\label{solurho222}
\ee
where now
\be
\alpha \in (-c_2,-c_1)\cup (c_1,c_2)\ ,\quad c_1=2\, {\rm arccot} (a)\ ,\ \ c_2=2\, {\rm arccot} (b)\ ,
\ee
\be
1+a^2=  \frac{4 \left(1+\tau +\sqrt{2 \tau +1}\right)}{\lambda  \tau ^2}\ ,
\ee
\be
1+b^2=  \frac{4 \left(1+\tau -\sqrt{2 \tau +1}\right)}{\lambda  \tau ^2}\ .
\ee
One can check  that this solves the saddle-point integral equation \eqref{ecurhoB}.
A plot of the density is shown in fig. \ref{eigentwo}. 

In the $\lambda\to 0$ limit, the cuts become small with a size scaling like $\sqrt{\lambda}$ and approach the origin.
The eigenvalue density approaches the scaling form
\be
\rho_{\rm 2\, cuts}(y)\bigg|_{\lambda\to 0} = \frac{1}{\pi\, |y|}\, \sqrt{y^2-c_-}\sqrt{c_+-y^2}\ ,\qquad y=\frac{\alpha}{\sqrt{\lambda }}\ ,
\ee
$$
c_\pm  =1+\tau\pm \sqrt{1+2\tau}\ .
$$

\begin{figure}[h!]
 \centering
 \includegraphics[width=0.45\textwidth]{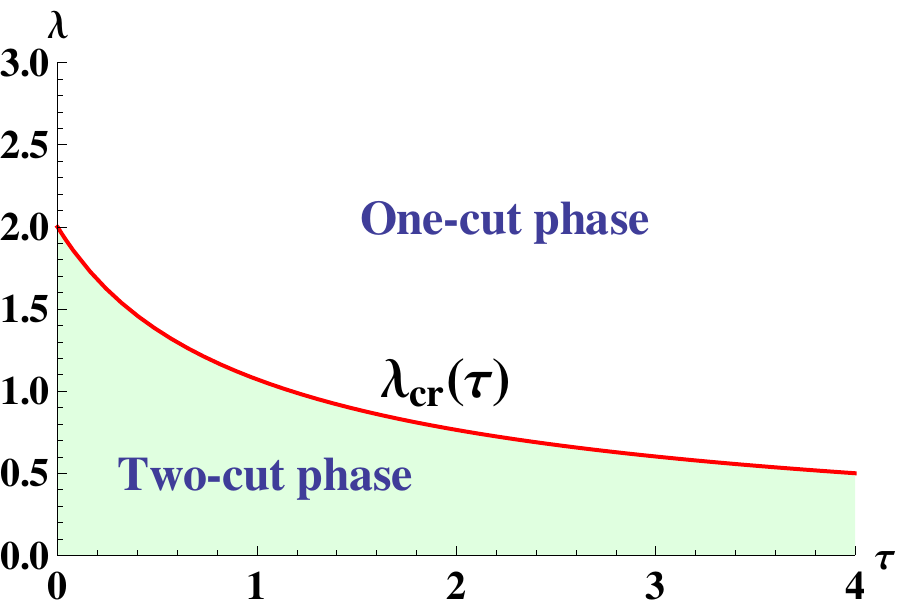}
 \caption{Phase diagram for model B. The critical line $\lambda_{\rm cr} $ separates
 a one-cut from  a two-cut phase. At $\tau=0$ it approaches the critical coupling
 $\lambda=2$ of the GWW matrix model.}
 \label{dosfases}
 \end{figure}

Summarizing, in the region $\tau>0,\ \lambda>0$, model B has two phases:
a strong coupling phase $\lambda>\lambda_{\rm cr}(\tau)$ described by a one-cut eigenvalue distribution and a weak coupling phase $\lambda<\lambda_{\rm cr}(\tau)$ described by a two-cut eigenvalue distribution. In the $\tau=0$ limit,
the eigenvalue distributions reduce to the ungapped ($\lambda>2)$ and gapped ($\lambda<2$) eigenvalue distributions of the GWW model.
The phase diagram of the theory is shown in fig. \ref{dosfases}.

\section{Critical behavior of the free energy and Wilson loops}

Let us now study the analytic properties of the free energy in crossing the critical line. To approach the critical line,
$\tau $ will be fixed and $\lambda$ will be increased or decreased.
The first derivative of the free energy is directly related to the vacuum
expectation value $W$ of the Wilson loop operator corresponding to a single plaquette,
$\frac{1}{2N}{\rm Tr}(U+U^ \dagger)$. One has 
\be
 W = \int_L d\alpha\ \rho(\alpha)\  \cos\alpha =\frac{\lambda^2}{2N^2}\frac{\partial F}{\partial \lambda}= \langle \cos\alpha\rangle \ .
\label{inteW}
\ee
The second and third derivatives of  the free energy can be obtained
by further differentiating the VEV of the Wilson loop operator.
That is, we will also need
\be
\partial_\lambda W\ ,\qquad \partial_\lambda^2 W \ .
\ee

The VEV of the Wilson loop  was computed in \cite{Russo:2020eif} in model A in the region $\lambda>0$, where the model is in the one-cut phase.
This calculation shows that the GWW  transition taking place at $\lambda=2$, $\tau=0$, becomes a crossover for any $\tau>0$.
As explained above, there is however a phase transition taking place on a critical line at negative $\lambda $.
For model B, this phase transition occurs in the $\lambda>0$ region.
Below we shall consider this model and compute the Wilson loop and its derivatives
in the two phases.

\subsection{Strong-coupling phase}

At strong coupling $\lambda>\lambda_{\rm cr}$, the density is given by 
\eqref{rhosenos}. 
To compute the integral \eqref{inteW}, it is convenient to use the density \eqref{soluno}
and then the map $\lambda\to -\lambda$, $\langle \cos\alpha \rangle\to -\langle \cos\alpha \rangle$, the latter induced by the shift in $\alpha$. An integration by residues then gives
\be
W= \frac{1}{\lambda}\, m(2-m)-\tau \left(1-\sqrt{1-m}\right)\ .
\ee
Here $m=m(\lambda,\tau)$ is obtained as one of the roots of the cubic equation
that arises from the normalization condition \eqref{normalsin} (we omit the explicit expression).
At small $\tau$, the Wilson loop in this phase has the expansion
\be
W=\frac{1}{\lambda}-\tau +\frac{\lambda\, \tau^2}{\lambda+2}+O(\tau^3)\ ,\ \ \ \lambda>\lambda_{\rm cr}\ .
\label{smalltau}
\ee
When $\tau\to 0$ this reproduces the expression for the Wilson loop in the  GWW matrix model in the ungapped phase.

One can compute derivatives of $W$ using the chain rule:
\be
\frac{dW}{d\lambda} =\frac{\partial W}{\partial \lambda}+\frac{\partial m}{\partial \lambda }\frac{\partial W}{\partial m}\ ,
\ee
etc., where $\frac{\partial m}{\partial \lambda }$ is obtained from the normalization condition \eqref{normalsin}.
In particular, we find the simple formula
$\frac{dW}{d\lambda}=-\frac{m^2}{\lambda^2}$.
At the critical point
\be
m\to m_{\rm cr} =1 - \frac{\lambda^2\tau^2}{16}\ .
\label{mcriti}
\ee
Using \eqref{lambdacriplus} and \eqref{mcriti}, we obtain the following exact expressions:
\bea
&& W(\lambda_{\rm cr},\tau)=\frac{1}{\tau^2}\left( (1+2\tau)^\frac32 -1-3\tau-\tau^2\right)\ ,
\label{Wcrita}\\ \nonumber \\
&& \frac{dW}{d\lambda}\bigg|_{\lambda_{\rm cr}}=
-\frac14 (1+2\tau)\ ,
\label{Wpcrita}\\ \nonumber\\
&&\frac{d^2W}{d\lambda^2}\bigg|_{\lambda_{\rm cr}}=
\frac{1}{8} (1+\tau)(1+2\tau)+\frac{1}{16} \left(2+4\tau+\tau^2\right)\sqrt{2 \tau +1}\ .
   \label{Wppcrita}
\eea

\subsection{Weak-coupling phase}

Let us now consider the two-cut phase described by the eigenvalue density \eqref{solurho222}.
Introducing the variable $t=e^{i\alpha}$ as in \textsection 2.1.2, 
the VEV of the Wilson loop can be expressed in terms of the following integral
\be
 W = \langle \cos\alpha \rangle =\int_L dt\,  \hat \rho(t) \, \frac{1+t^2}{2t}\ .
\ee
In order to compute the integral, we use the shifted density \eqref{solurho2} as
before, taking into account the map $W\to -W$ and $\lambda\to -\lambda$. We find the following remarkably simple result:
\be
 W  = 1 - \frac{\lambda}{4}(1+2\tau)\ ,\qquad \qquad 0<\lambda <\lambda_{\rm cr}\ .
\ee
For $\tau\to 0$, it reproduces the Wilson loop of the GWW matrix model in the gapped phase.

\begin{figure}[h!]
 \centering
 \includegraphics[width=0.45\textwidth]{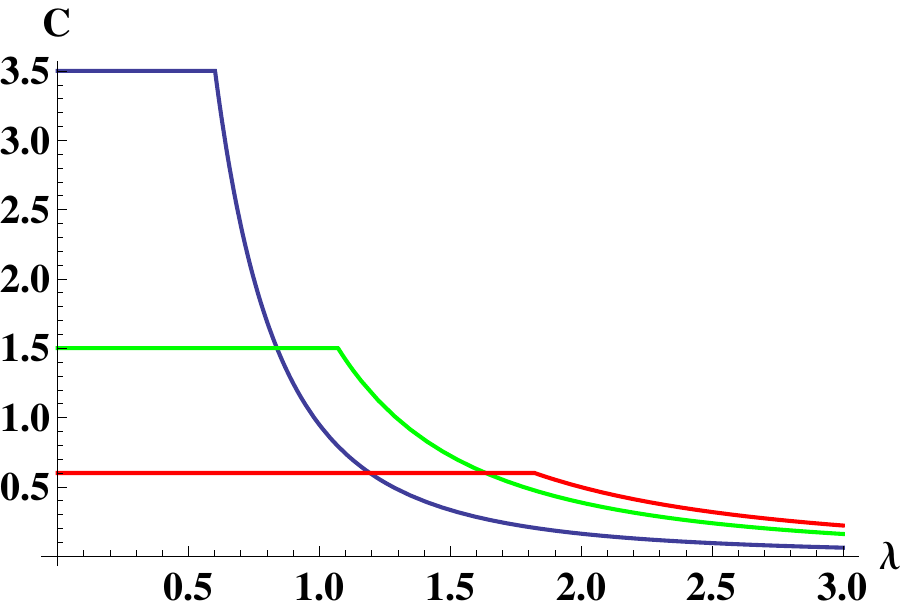}
 \caption{The specific heat $C/N^2=-2\partial_\lambda W$ vs. $\lambda $, for
  $\tau=0.1$ (red), $\tau=1$ (green) and $\tau=3$ (blue). The discontinuity in the
  derivative at the critical point shows that the  phase transition is third-order.}
 \label{especifico}
 \end{figure}

On the critical line $\lambda_{\rm cr}(\tau ) $, $W$ reduces to the same expression
\eqref{Wcrita} of the supercritical phase, so the Wilson loop is continuous
across the critical line.
Similarly, the first derivative of the Wilson loop matches \eqref{Wpcrita}.
For the second derivative, we now obtain
\be
\frac{d^2  W}{d\lambda^2}\bigg|_{\lambda_{\rm cr}}\equiv 0\ .
\ee
Comparing with \eqref{Wppcrita}, one finds that the second derivative of the Wilson loop is discontinuous across the transition.
This implies that the third-derivative of the free energy is discontinuous. Thus,  the system undergoes a third-order phase transition across the critical line
represented by $\lambda =\lambda_{\rm cr}(\tau)$. Figure \ref{especifico} displays the behavior of the first derivative $\partial_\lambda W$, related to the specific heat (viewing the system as a statistical ensemble with temperature $T=\lambda $).

\subsection{Winding Wilson loops}

In a similar way one can compute the vacuum expectation value of winding Wilson loops~\cite{Gross:1980he} (see also \cite{Okuyama:2017pil,Rossi:1996hs,Alfinito:2017hsh}). At large $N$, they are given by
\be
W_k= \frac{1}{2N} \, \langle {\rm Tr}\, \left( U^k+{U^\dagger}^k\right)\rangle=\int_L d\alpha\ \rho(\alpha)\  \cos (k\alpha) = \langle \cos(k \alpha)\rangle\ .
\ee
We obtain the following results for the two phases:

\subsubsection*{Strong coupling phase $\lambda>\lambda_{\rm cr}$}

\bea
W_2 &=& -\frac{2 (1-m)^2 m}{\lambda }+ \left( (1+m)\sqrt{1-m} -1\right) \tau  \ ,
\nonumber\\
W_3 &=& \frac{m (1-m)^2  (2-5 m)}{\lambda }+\left( (2m^2+1) \sqrt{1-m} -1\right) \tau   \ ,
\label{wikstrong}
\eea
etc.
At small $\tau $, they have the expansion
\bea
W_k &=& -\tau +\frac{k\lambda}{2+\lambda}\, \tau^2  +O(\tau^3)\ ,\qquad \qquad k\geq 2\ .
\eea
The $W_k$'s vanish in the limit $\tau \to 0$, in agreement with \cite{Gross:1980he}.

\subsubsection*{Weak coupling phase $\lambda<\lambda_{\rm cr}$}

\bea
W_2 &=& 1-\lambda  (2 \tau +1)+
\frac{1}{4} \lambda ^2 \left(2 \tau ^2+3 \tau +1\right) \ ,
\nonumber\\
W_3 &=& 1  -\frac{9}{4} \lambda  (2 \tau +1) +\frac{3}{2}
   \lambda ^2 \left(2 \tau ^2+3 \tau +1\right)
-\frac{1}{16} \lambda ^3 \left(8 \tau ^3+24 \tau ^2+20 \tau +5\right)\ ,
\label{wikweak}
\eea
etc. One can check that \eqref{wikstrong} and \eqref{wikweak} match on the critical line $\lambda=\lambda_{\rm cr}(\tau)$, whereas the second derivatives of $W_k$ 
are discontinuous.

\section{The $\beta$ function}

One of the interesting features of the GWW matrix model is that a $\beta$ function
can be constructed for the large $N$ theory through the dependence
of the effective 't Hooft coupling on the lattice spacing \cite{Gross:1980he,Wadia:2012fr}.
The  properties of the $1/N$ expansion of the  $\beta $ function have been recently discussed in  \cite{Ahmed:2018gbt}. It turns out that
the complete non-perturbative trans-series can be described thanks to the fact that the $\beta$ function can be expressed in terms of the VEV of $\det U$, which satisfies a differential equation for any value of $N$. 
We expect that a similar treatment can be carried out in the deformed GWW model. In this section we will just focus in the leading large $N$ $\beta$ function for the models A and B.

In terms of the string tension $\sigma $ and the lattice spacing $a$, the Wilson loop has the form
\be
W(\lambda,\tau)=e^{-a^2\sigma }\ .
\ee
Following \cite{Gross:1980he,Wadia:2012fr}, a running coupling $\lambda(a) $ can be obtained by varying the lattice spacing $a$ keeping $\sigma $ fixed. 
This defines an effective coupling $\lambda =\lambda(a;\tau)$. 
We assume that $\tau $ does not renormalize as here
it represents the Veneziano parameter $N_f/N$.
The $\beta$ function for the coupling $\lambda$ is then obtained by the formula
\be
\beta  = - a\frac{\partial \lambda}{\partial a}\ .
\ee
This leads to the following expression in terms of the Wilson loop:
\be
\beta  = - \frac{2W}{\partial_\lambda W} \, \ln W\ .
\label{formulita}
\ee

\subsubsection*{Model A}

\begin{figure}[h!]
 \centering
 \includegraphics[width=0.45\textwidth]{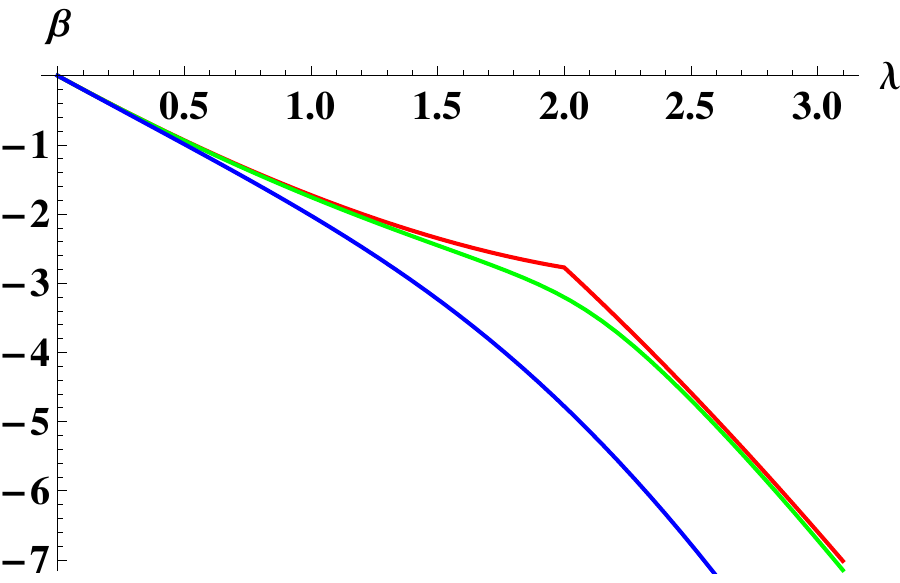}
 \caption{The $\beta$ function for model A, for $\tau=0$ (red), $\tau=0.03$ (green), $\tau=0.35$ (blue).}
 \label{betacos}
 \end{figure}

In the region $\{ \lambda>0,\, \tau>0\}$,  model A has only one phase. The Wilson loop has been calculated in \cite{Russo:2020eif} and can be obtained from the formulas in
\textsection 3.1 by the map $\lambda\to -\lambda$, $W\to -W$.
The ensuing $\beta $ function \eqref{formulita} is plotted in fig. \ref{betacos} for different values of  $\tau $. One can see that the discontinuous behavior of the $\tau=0$ GWW 
matrix theory is smoothed out for finite positive $\tau$. 

The  $\beta $ function has the following perturbative (planar) expansion
\be
\beta =-2 \lambda +\frac{1}{4}(1-2 \tau )\ \lambda ^2 
+\frac{1}{48}(1-12 \tau ) \lambda ^3 +O\left(\lambda ^4\right)\ ,\qquad 0<\lambda\ll 1\ ,
\ee
where the first term comes from the classical dimension of the coupling.

\subsubsection*{Model B}

Model B has two phases in the region $\{\lambda>0,\, \tau>0 \}$. 
In the weak coupling phase, one can use the results of
 \textsection 3.2 to derive the following simple formula 
\be
\beta=\frac{2}{1+2\tau}
\left( 4-\lambda(1+2\tau) \right)
\ln \left(1-\frac14 \lambda(1+2\tau) \right)\ ,\qquad \lambda<\lambda_{\rm cr}(\tau)\ .
\ee
This has the perturbative expansion
\be
\beta =-2 \lambda +\frac{1}{4} \lambda ^2 (2 \tau +1)+\frac{1}{48} \lambda ^3 (2 \tau +1)^2+O\left(\lambda ^4\right)\ .
\ee

\begin{figure}[h!]
 \centering
 \begin{tabular}{cc}
 \includegraphics[width=0.4\textwidth]{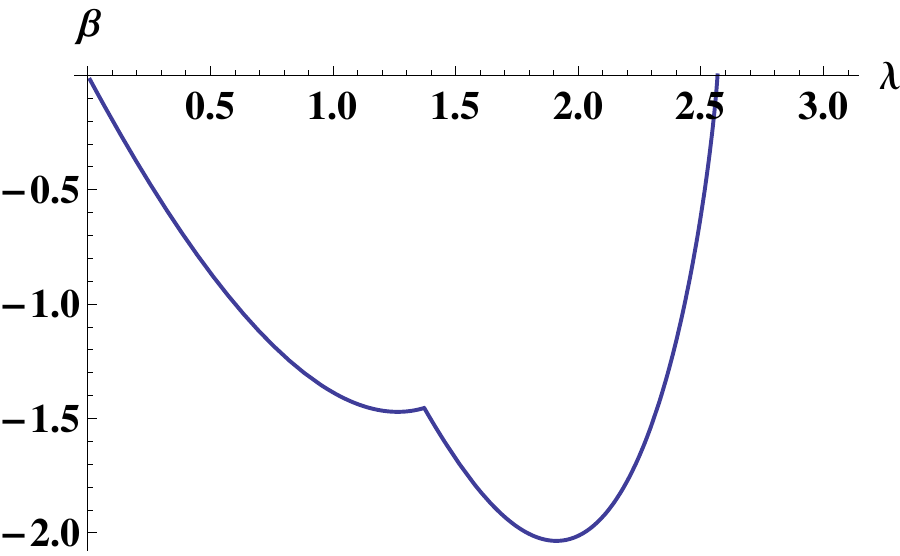}
 &
 \qquad \includegraphics[width=0.4\textwidth]{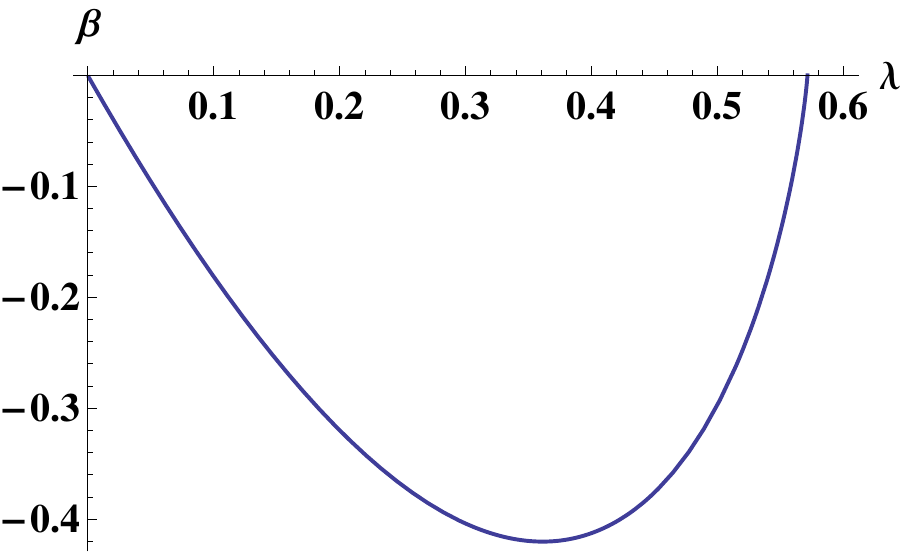}
 \\ (a)&(b)
 \end{tabular}
 \caption
 {The $\beta $ function for model B. (a) $\tau=0.5$ and (b) $\tau=3$. In case (b) the fixed point is reached before the critical coupling $\lambda_{\rm cr}$.}
 \label{betatwofase}
 \end{figure}

In the strong coupling phase the resulting
expression is obtained from \eqref{formulita}, where $W$ and $\partial_\lambda W$
can be read from the formulas in \textsection 3.1. In either phases, for $\tau =0$, one reproduces the $\beta $ function found by Gross and Witten \cite{Gross:1980he}.

The $\beta$ function is shown in fig. \ref{betatwofase} for different values of $\tau$. Comparing with  the $\tau=0$ case, a new feature appears for any $\tau>0$.
The $\beta $ function now has an IR stable fixed point at finite coupling $\lambda $.
Coming from $\lambda=0$, when $\tau <1+\sqrt{2}$, the fixed point occurs after the phase transition, that is,  in the
one-cut (strong coupling) phase.
When $\tau >1+\sqrt{2}$, the fixed point occurs before the phase transition has taken place.
For small $\tau$, the fixed point occurs at $\lambda^*\approx 1/\tau $, {\it i.e.} at  strong coupling $\lambda^*\gg 1$ . This can be seen from the expansion \eqref{smalltau}. 
It is important to note that the origin of the fixed point is that $W\to 0$, which is indicative of an infinite string tension, a highly confining regime. This is to be distinguished from a fixed point where
$W\to 1$, which would  indicate a vanishing string tension and a non-confining regime, which does not happen in this case (and it is not expected to happen in general in two dimensions).


\subsection*{Acknowledgments}
The author is grateful to M. Tierz and  K. Zarembo for valuable comments.
We acknowledge financial support from projects 2017-SGR-929, MINECO
grants FPA2016-76005-C and  PID2019-105614GB-C21.

\end{document}